\documentclass[10pt,letterpaper,twocolumn]{article} 

\usepackage{ol2}
\usepackage[draft]{hyperref}

\usepackage{graphicx}
\usepackage{amssymb,amsmath}
\renewcommand{\d}[1][{\negthickspace}]{\mathrm{d}{#1}\;}

\begin{document}
\bibliographystyle{osajnl}
\twocolumn[ 
\title{Beyond the frame rate: Measuring high-frequency fluctuations with light intensity modulation}
\author{Wesley P. Wong$^*$ and Ken Halvorsen$^\dagger$}
\address{The Rowland Institute at Harvard, Harvard University, Cambridge, Massachusetts 02142, USA
\\
$^*$Corresponding author: wong@rowland.harvard.edu \ \ 
$^\dagger$Both authors contributed equally
\\
Journal-ref: Optics Letters 34(3), 277--279 (2009) }




\begin{abstract}

Power spectral density measurements of any sampled signal are typically restricted by both acquisition rate and frequency response limitations of instruments, which can be particularly prohibitive for video-based measurements. We have developed a new method called Intensity Modulation Spectral Analysis (IMSA) that circumvents these limitations, dramatically extending the effective detection bandwidth. We demonstrate this by video-tracking an optically-trapped microsphere while oscillating an LED illumination source. This approach allows us to quantify fluctuations of the microsphere at frequencies over 10 times higher than the Nyquist frequency, mimicking a significantly higher frame rate.
  
\end{abstract}



] 



Measuring the power spectral density (PSD) is a useful way to
characterize fluctuations and noise for a diverse range of physical
processes. Optical measurements of the PSD have been used to study single molecule dynamics \cite
{oddershede2002msm}, bacterial chemotaxis and motion
\cite{gabel2003sfr,korobkova2004mnb}, quantum dot blinking \cite
{pelton2004cqd} and 
microrheology \cite{gittes1997mvs, valentine2001imi}, and to calibrate optical traps \cite {svoboda1994bao}. Unfortunately, limited acquisition rate and detector frequency response restrict PSD measurements in frequency space, which is especially detrimental to video applications.
%
Notably, the highest frequency that can be sampled directly, with few exceptions \cite {verdun1988bnl, elshafey2004bnl, wong2006eit, donciu2005bnc}, is half the acquisition rate, or Nyquist frequency.


We have developed a method called Intensity Modulation Spectral
Analysis (IMSA), which overcomes
these limitations in a simple and economical way. By simply oscillating the light intensity of an optical signal prior to detection, the PSD
at the oscillation frequency can be determined from the measured variance, even if that frequency is above the acquisition rate. This is similar in spirit to signal processing methods that can spectrally shift a signal (e.g. heterodyne detection, lock-in techniques), or extract high-frequency information folded down via aliasing (e.g. undersampling \cite{donciu2005bnc, vaughan1991tbs}). Practically, IMSA can dramatically extend the frequency range of an existing measurement device, allowing, for example, an inexpensive camera to be used in place of a significantly more expensive one. Here we present the framework for IMSA and an
experimental demonstration using an optically-trapped microsphere, in which modulating the brightness of an LED allows PSD measurement well beyond the Nyquist frequency and camera frame rate.





\paragraph{\textbf{IMSA: Concept and foundations}}
Physical acquisition systems do not make instantaneous measurements when sampling a signal,
but rather collect data over finite integration times.  Consider a
stationary random trajectory $X(t)$.
The measured
trajectory $X_m(t)$ can be expressed as a convolution of the true trajectory and an impulse
response $H(t)$:
\begin{equation} \label{eq:conv}
X_{m}(t)= (X \ast H)(t) \equiv \int X(t')H(t-t') \d[t']
\end{equation}
Then the measured power spectrum $P_m(\omega)$ differs from
$P(\omega)$, the true power spectrum of $X$, according to the relation
$P_m (\omega ) =P(\omega) \vert \tilde{H}(\omega ) \vert^2$, and the
total measured variance is the integral of $P_m (\omega )$
over all frequencies:
\begin{equation}\label{eq:measuredvarfourier}
  \text{var}[X_m]=\frac{1}{2\pi}\int P(\omega )
  \left\vert \tilde{H}(\omega ) \right\vert^2
  \d[\omega]
\end{equation}
where the tilde designates the Fourier transform, $\tilde{X}(\omega) =
\int X(t) \exp(-i \omega t)\d[t]$. Integrals are taken from $-\infty$
to $+\infty$ unless otherwise specified.

In the simplest case, the measured quantity $X_m$ is the unweighted time average of the true value $X$ over the integration time $W$, i.e. the impulse response is a rectangular
function:
\begin{equation} \label{eq:filter}
H_0(t)=\left\{ \begin{array}{cl}
\frac{1}{W} & -W/2 < t \le W/2 \\
0 & \text{elsewhere}
\end{array}\right.
\end{equation}
Correspondingly, the measured power spectrum is the original power
spectrum multiplied by:
\begin{equation} \label{eq:powerspech}
\left\vert \tilde{H}_0(\omega) \right\vert^2= \left(
	   {\frac{\sin(\omega W/2)}{\omega W/2}} \right)^2
\end{equation}

This simple case of a rectangular impulse response is a good model for
video-imaging acquisition systems, where $W$ is the
exposure time. This averaging leads to the common problem of video
image blur, which not only adds errors in the position of tracked
objects, but also causes systematic biases when quantifying
fluctuations \cite {yasuda1996dmt, savin2005sad, savin2005rfe,
  wong2006eit}. As we have previously demonstrated \cite
{wong2006eit}, by measuring the variance for different exposure times
$W$ and fitting to equation \ref{eq:measuredvarfourier}, the power
spectrum can be characterized above the acquisition rate of the
detection system.  However, this approach requires that the functional form of
the power spectrum is known a priori.

Interestingly, by oscillating the intensity of a source signal (e.g. light
in the case of video imaging) and measuring the variance of the
resulting signal, the power spectrum can be reconstructed without
prior knowledge.  This is the fundamental idea behind IMSA.

During detection, the finite integration time $W$ causes the true
power spectrum to be multiplied by the low-pass filter $\vert
\tilde{H}_0(\omega) \vert^2$ (equation \ref{eq:powerspech}). As $W$
becomes longer, $\vert \tilde{H}_0(\omega) \vert^2$ approaches an
unnormalized delta function, i.e. $\vert \tilde{H}_0(\omega) \vert^2
\rightarrow \alpha \delta(\omega)$. Multiplying the rectangular
impulse response by a complex exponential shifts this delta function,
i.e. if $H(t) = \exp(-i\omega' t) H_0(t)$, then $\vert
\tilde{H}(\omega) \vert^2 \rightarrow \alpha \delta(\omega -
\omega')$. From equation \ref{eq:measuredvarfourier} we see that the
variance approaches $\alpha P(\omega') / 2\pi$.
Thus, as demonstrated in this simple example, the power spectrum can
be sampled by shifting the filter to any frequency of interest
$\omega'$ and measuring the variance.

\begin{figure}[tbp]
  \begin{center}
    \includegraphics[width=0.48\textwidth]
    {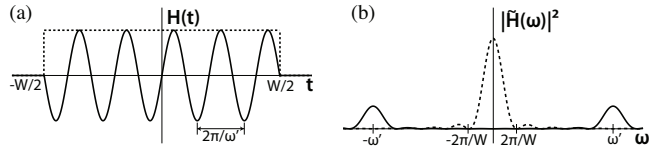}
    \caption{\label{fig:sinc_figure}Graphical representation of IMSA showing (a) unnormalized impulse responses $H(t) = L(t) H_0(t)$ for $\phi=0$ and $B = 0$ (solid line, see equation \ref{eq:realimpulseresponse})  and $H(t) = H_0(t)$ (dotted line, see equation \ref{eq:filter}), and (b) the squared magnitude of their Fourier transforms, which filter $P(\omega)$ in terms t1 and t2 of equation \ref{eq:measuredvarfullexpressioneven}.}
  \end{center}
\end{figure}

Practically, we use the real-valued impulse response
\begin{equation}\label{eq:realimpulseresponse}
H(t) = L(t) H_0(t)/N
\end{equation}
where $L(t) \equiv \sin(\omega' t + \phi) + B$ 
and $N$ is a normalization factor (see Fig. \ref{fig:sinc_figure}). We set $N = \int L(t) H_0(t) \d[t]$
so that the convolution represents a time-averaged signal weighted by
the source intensity modulation $L(t)$, and let $B\ge0$; the parameters ($W$, $B$)
should be chosen such that $N \ne 0$. Note that $N=B$ whenever $\phi$
(the phase-shift between modulation and sampling) is an integer
multiple of $2\pi$, or whenever there are an integer number of
oscillations within each exposure window.

By substituting $\tilde{H}(\omega)$ (the Fourier transform of equation \ref{eq:realimpulseresponse}) into equation \ref{eq:measuredvarfourier}, and
noting that $P(-\omega) = P(\omega)$ for a real-valued wide-sense
stationary process, we obtain for the measured variance:
%
\begin{align} 
 & \text{var}[X_m] =
  \frac{1}{4 \pi N^2} \int P(\omega) \left\vert \tilde{H}_0(\omega -
    \omega') \right\vert^2 \d[\omega] \tag{\emph{t1}} \\
  & \quad + \frac{B^2}{2 \pi N^2} \int P(\omega ) \left\vert \tilde{H}_0(\omega)
  \right\vert^2 \d[\omega] \tag{\emph{t2}}\\
  & \quad + \frac{B \sin(\phi)}{\pi N^2} \int P(\omega) \tilde{H}_0(\omega)
  \tilde{H}_0(\omega-\omega') \d[\omega] \tag{\emph{t3}} \\
  & \quad - \frac{\cos(2\phi)}{4 \pi N^2} \int P(\omega) \tilde{H}_0(\omega+\omega')
  \tilde{H}_0(\omega-\omega') \d[\omega] \tag{\emph{t4}} \\
  \label{eq:measuredvarfullexpressioneven}
\end{align}
This key equation can be used to determine the power spectrum from the
measured variance, since term \emph{t1} approaches $P(\omega')/2WN^2$
as $W$ gets longer (see
endnote \cite{endnote1} and Fig. \ref{fig:sinc_figure}), while terms \emph{t3} and \emph{t4} become small and term \emph{t2}
can be measured directly.  Additionally, proper phase selection can completely eliminate \emph{t3} and \emph{t4}.

\paragraph{\textbf{IMSA Usage}}
To determine the power spectral density of a signal at the frequency
of interest $\omega'$, the signal should be convolved with $L(t) H_0(t)/N$ (equation \ref{eq:realimpulseresponse}) by modulating its intensity, and then its variance should be
measured.  Writing the measured variance as
$\text{var}[X_m](\omega')$, the PSD is given by the
following formulas:
\\
When $B=0$,
\begin{equation}\label{eq:simbasummary1}
  P(\omega') = 2 W N^2\text{var}[X_m](\omega')
\end{equation}
provided $\phi = \pi/4 + m\pi/2$, where $m$ is an integer.
\\
When $B \ne 0$,
\begin{equation}\label{eq:simbasummary2}
P(\omega') = 2 W B^2 \left( \text{var}[X_m](\omega') -
  \text{var}[X_m](0) \right)
\end{equation}
provided that the exposure time $W$ is chosen such that $W = n 2 \pi /
\omega'$ where $n$ is a natural number, and $\phi$ is chosen according
to (see endnote \cite{endnote2}):
\begin{equation}\label{eq:phaseformula3}
\phi = (-1)^{n} \sin^{-1}\left(\sqrt{B^2 + 1/2} - B \right)
\end{equation}

On a practical note, terms \emph{t3} and \emph{t4} are small when $W \gg 1/\omega'$ (i.e. there are many oscillation cycles within each exposure window).  If both cross-terms \emph{t3} and \emph{t4} are
negligible, the selection of $\phi$ is irrelevant. In fact, oscillations may not even have to be synchronized to the acquisition device to make a good IMSA measurement---simply
multiplying the input signal by $L(t)$ prior to detection can yield
acceptable results.

The error in $P(\omega')$ is governed primarily by the error in the
variance (e.g. for $N$ samples the relative statistical standard error
is $\sqrt(2/N)$, barring instrumental error \cite{
  wong2006eit}). The resolution in $\omega'$ is given by the width of
$\vert \tilde{H}_0(\omega) \vert^2$ (see Fig.
\ref{fig:sinc_figure}), which is approximately $\pi/W$ in each
direction (77\% of the area under the curve).

\paragraph{\textbf{Experimental Demonstration}}

\begin{figure}[tbp]
  \begin{center}
    \includegraphics[width=0.4\textwidth]
    {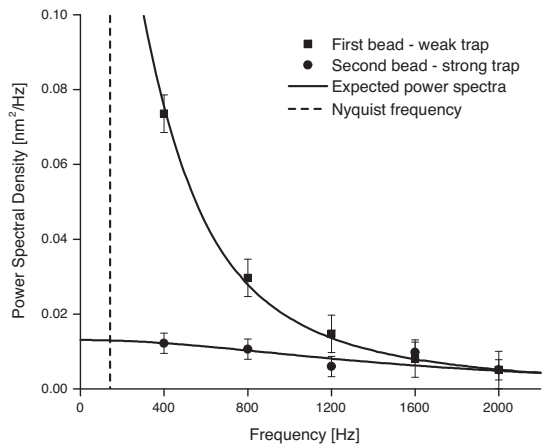}
    \caption{\label{fig:simba_data} Measurement of the power spectrum
      using IMSA for two trapped beads with different spring constants
      (data points). The expected power spectra are superimposed
      (solid lines), showing good agreement well beyond the Nyquist
      frequency (vertical dashed line). Error bars represent
      statistical error.} 
  \end{center}
\end{figure}

To demonstrate IMSA experimentally, we measured the power spectrum of
an optically trapped polystyrene microsphere (2.5 um, Corpuscular)
using a machine vision camera (GE680, Prosilica) and an LED (MRMLED,
Thorlabs) with intensity modulation capable of up to 300kHz (custom
current source, Rowland Institute electronics lab).  Details of the
particle tracking have been discussed previously for a functionally
identical setup, where it was demonstrated that the system is
well-described by equation \ref{eq:filter}
\cite{wong2006eit}. The light intensity was sinusoidally modulated (as in equation \ref{eq:realimpulseresponse}) with $B=0.5$ and $\phi$ determined
from equation \ref{eq:phaseformula3}, and the integration time of the
camera was $W=2.5$ ms. Modulated light at 5 different frequencies
$\omega'$ was interspersed with DC light on a frame by frame basis.
Each of the 10 variances was calculated and each PSD data point was
calculated according to equation \ref{eq:simbasummary2}, where the
mean of the two neighboring DC frames was used as $
\text{var}[X_m](0)$ for each of the 5 measurements.  Data was
collected for two beads at different laser powers. The spring constant
and friction factor for each bead was measured using the
blur-corrected power spectrum method \cite{wong2006eit}.  The IMSA measured values, which have no
free fitting parameters, are in excellent agreement with the expected
power spectra (Fig. \ref{fig:simba_data}).

\paragraph{\textbf{Concluding Discussion}}

As we have shown, IMSA provides a unique and practical way for
measuring the PSD of a signal, which overcomes acquisition rate and frequency
response limitations of instruments.  While especially beneficial to video imaging applications (due to the almost negligible cost of implementation compared with the expense of fast video cameras),  IMSA is a general method applicable to signals in which the time-averaged weighting can be controlled before sampling.

Owing to its potential to benefit a broad range of fields and its flexibility of implementation (see endnote \cite{endnote3}), we expect that IMSA will become a common laboratory method for measuring power spectra. Microrheology measurements
\cite{crocker1996mdv, valentine2001imi} can take immediate
advantage of the up to $\sim 10 000$ fold increase in frequency range
over standard video imaging, without losing the ability to track multiple targets. Widely used fluorescence-imaging
techniques \cite {toprak2007nft} stand to benefit as well, with IMSA enabling measurements at frequencies which are currently impossible by any other method owing to light limitations.

The dramatic increase in frequency range enabled by IMSA can be used to push the
envelope of high frequency measurement and to realize significant cost
savings in instrumentation. Using an inexpensive
LED illuminator, we transformed a standard video camera into a
spectrum analyzer with a frequency range of up to $\sim 300$ kHz.



\paragraph{\textbf{Acknowledgments}}
Thanks to M.~M.~Burns,
  W.~Hill, A.~M.~Turner, and M.~M.~Forbes for helpful conversations,
  and C.~Stokes for the LED
  current source. Funding was provided by the Rowland Junior Fellows
  program.
\end{document}